%% file: main.tex
\documentclass[a4paper,11pt]{article}
\pdfoutput=1 

\usepackage{style/jinstpub} 
\usepackage{pdflscape}
\usepackage{afterpage}
\usepackage{adjustbox,lipsum}
\usepackage{graphicx}
\usepackage{dcolumn}
\usepackage{bm}
\usepackage{amsmath}
\usepackage{diagbox}
\usepackage{rotating}
\usepackage{multirow}
\usepackage{xcolor}
\usepackage{hepunits}
\usepackage{multirow}
\usepackage{color}
\usepackage{hyperref}
\usepackage[mathlines]{lineno}
\usepackage{xspace} 
\usepackage{verbatim} 
\usepackage{listings}
\usepackage{lineno}

\newcommand{\geant}{\textsc{Geant4}\xspace}

\title{The DAMIC-M Low Background Chamber}
\input{0_authorlist.tex}

\abstract{The DArk Matter In CCDs at Modane (DAMIC-M) experiment is designed to search for light dark matter (m$_{\chi}$<10\,GeV/c$^2$) at the Laboratoire Souterrain de Modane (LSM) in France. DAMIC-M will use skipper charge-coupled devices (CCDs) as a kg-scale active detector target. Its single-electron resolution will enable eV-scale energy thresholds and thus world-leading sensitivity to a range of hidden sector dark matter candidates. A DAMIC-M prototype, the Low Background Chamber (LBC), has been taking data at LSM since 2022. The LBC provides a low-background environment, which has been used to characterize skipper CCDs, study dark current, and measure radiopurity of materials planned for DAMIC-M. It also allows testing of various subsystems like readout electronics, data acquisition software, and slow control. This paper describes the technical design and performance of the LBC.}


\begin{document}
\maketitle

\section{Introduction}
A multitude of astrophysical and cosmological observations have revealed that our universe is mostly made of non-luminous, non-baryonic dark matter (DM)\,\cite{Zwicky:1933,Rubin:1970,Ostriker:1974}. Measurements of 
type Ia supernovae\,\cite{SupernovaSearchTeam:1998fmf,SupernovaCosmologyProject:1998vns}, cosmic microwave background fluctuations\,\cite{Planck:2018vyg}, the large-scale distribution of galaxies\,\cite{2dFGRS:2005yhx,SDSS:2006lmn,BOSS:2016wmc,eBOSS:2020yzd,DES:2021wwk}, weak gravitational lensing\,\cite{Schrabback:2009ba,Heymans:2013fya,Jee:2015jta,HSC:2018mrq,Heymans:2020gsg}, and the abundances of the light elements (see a review in Ref.\,\cite{ParticleDataGroup:2022pth})
suggest that the Standard Model of Cosmology, $\Lambda$CDM, remarkably describes the total matter density when parameterized with a cold DM particle. There are many particle candidates that fit the constrained properties, yet no experiment has seen evidence for direct interactions between DM and the Standard Model particles. 

Over the last few decades, Weakly Interacting Massive Particles (WIMPs)\,\cite{Steigman_Turner} have been the most searched-for DM candidate. With a mass larger than the proton ($\approx 1$~GeV$/c^2$), these particles are highly motivated since if their annihilation cross section is similar to the electroweak-scale, it results in a thermal relic from the early universe consistent with the observed DM abundance. As multi-ton xenon experiments have only produced null results for DM masses larger than 10 GeV\,\cite{PandaX-4T:2021bab,LZ:2022lsv,XENON:2023cxc}, the field has recently expanded its mass range of interest to include light (sub-GeV) DM\,\cite{Essig:2022dfa}. Candidates in this regime include light WIMPs and dark or hidden sector particles\,\cite{Battaglieri:2017aum}. Due to the smaller mass of these particles, the kinematics favor lower energy recoils that require detectors to be sensitive to sub-keV energy deposits.

The DAMIC-M (DArk Matter In CCDs at Modane) experiment\,\cite{DAMIC-M:2023gxo} has been specifically designed to detect both nuclear and electronic recoils from sub-GeV DM scattering. The detector will operate deep underground at the Laboratoire Souterrain de Modane (LSM) and contain up to 208\,charge-coupled devices (CCDs) for a total sensitive mass of silicon of $\sim$700\,g. The CCDs feature skipper readout amplifiers allowing for the detection of DM-induced ionization events within the silicon with sub-electron resolution through non-destructive, repeated pixel measurements\,\cite{skipper,Chandler1990zz,Tiffenberg:2017aac,DAMIC-M:2022xtp}. The ability to detect single electrons with low dark current sensors\,\cite{DAMIC:2019dcn,SENSEI:2020dpa} in a very low background environment (<1\,dru\,\footnote{1\,dru = 1\,event/keV/kg/day.}), will allow DAMIC-M to lower the energy threshold down to only of a few\,eV. 

Here we report the technical details of the DAMIC-M prototype detector, the Low Background Chamber (LBC), which is currently in operation at LSM. The objectives of the detector are discussed in Sec.\,\ref{sec:objectives}, followed by the description of the detector design and instrumentation in Sec.\,\ref{sec:design} and \ref{sec:instruments}. The efforts to reduce the radioactive background and first results on detector performance are shown in Sec.\,\ref{sec:backgrounds} and \ref{sec:performance_results}, respectively.

\section{Prototype Objectives}
\label{sec:objectives}
The LBC was built to establish DAMIC-M infrastructure at LSM, test components and subsystems, and benchmark skipper CCD performance in a low background, underground environment. The detector has also served as a scientific instrument, producing preliminary background studies (see Sec.\,\ref{sec:performance_results} for more details) and first DAMIC-M dark matter-electron scattering exclusion limits\,\cite{DAMIC-M:2023gxo,DAMIC-M:2023hgj} 

LSM is the deepest underground laboratory in Europe, sitting 1700\,m below the Fr\'{e}jus Peak (4800\,m.w.e.) between the French and Italian Alps. In this location, the muon flux is reduced to 5.4\,$\mu$/m$^2$/day\,\cite{Piquemal:2012fs,EDELWEISS:2013kzp}, i.e., by more than a factor of 10$^6$ compared to sea level. Although several experiments have previously built facilities at LSM, such as the EDELWEISS dark matter\,\cite{EDELWEISS:2017lvq} and the NEMO/SuperNEMO neutrinoless double beta decay\,\cite{Arnold:2004xq,SuperNEMO:2010wnd} experiments, the DAMIC-M LBC is the first CCD detector to operate at this site. Thus, one of the major objectives of the prototype was for the collaboration to gain experience working with the LSM staff and begin building the cleanroom and supporting infrastructure for DAMIC-M. An ISO\,5 class cleanroom was built to house the LBC and will later be used for the full DAMIC-M detector. More details about the infrastructure can be found in Sec.\,\ref{sec:design_infrastructure}.

An important goal of the LBC was to demonstrate that a background rate below DAMIC at SNOLAB levels\,\cite{DAMIC:2021crr} i.e., <10\,dru  could be achieved at the LSM depth with cleaner detector components. Extensive R\&D work was required to select the most pure materials, determine procedures to properly clean surfaces, and to perform radioassays. Both experience from other low background experiments\,\cite{EDELWEISS:2013wrh, MAJORANA:2016lsk, XENON:2019izt, DAMIC:2021crr} and a custom \geant simulation\,\cite{GEANT4:2002zbu, Allison:2006ve, Allison:2016lfl} were used to inform these decisions, as further described in Sec.\,\ref{sec:backgrounds}. In addition, the LBC also acts as a test bed to validate the operation of detector subsystems for DAMIC-M (see Sec.\,\ref{sec:instruments}), such as the cryogenics, CCD electronics, slow control, and data quality monitoring. 

As the first opportunity for DAMIC-M to bring skipper CCDs underground, a crucial objective of the LBC was to evaluate the performance of these novel devices. Data were taken in different shielding configurations to examine the decrease of radiation background, measure the dark current induced by external radiation and select operating parameters and readout strategies depending on the background radiation level. Furthermore, observables -- such as gain, noise, energy resolution, dark counts -- were optimized and their variation over time was studied. We checked the production and packaging yield of large-format CCDs, observed the CCD performance after multiple cooling and warming cycles, and measured the dark current and clock-induced charge in low background environment. Sec.\,\ref{sec:performance_results} gives a summary of the various datasets and performance outcomes.

Dependent on the successful achievement of goals, the LBC was also designed to produce science results, because the single-electron resolution and low dark current of skipper CCDs makes them an ideal target and detector for light dark matter. The sensitivity of the LBC probes both models of thermal freeze-out and freeze-in\,\cite{Battaglieri:2017aum} for numerous dark matter masses within a few months of exposure. The low energy threshold of skipper CCDs also makes them  uniquely suited to search for mediators of the hidden sector.

First results on dark matter-electron scattering\,\cite{DAMIC-M:2022aks,DAMIC-M:2023gxo}, including a search for a daily modulation in the event rate\,\cite{Privitera:2024tpq,DAMIC-M:2023hgj}, were published, and studies of alternative low-mass dark matter candidates are forthcoming. In addition, three doctoral theses describe work on the LBC \cite{Traina:2022,DeDominics:2022,Piers:2023}.

The LBC went through various configurations and two of them are highlighted in this paper, particularly when we describe the detector design in Sec.\,\ref{sec:design} and achieved improvement of the radioactive background in Sec.\,\ref{sec:backgrounds}. Our adopted nomenclature for these two setups is as follows: The first one is Setup\,1 and it includes two CCDs, each of the pixel format 6k$\times$4k, placed in the CCD box fully made from oxygen-free high-conductivity (OFHC) copper. Setup\,2 consists of two CCD modules, each with four 6k$\times$1.5k-pixel format CCDs and one flex cable, in the CCD box with lids made from electro-formed copper (EFC). Let us stress that the two setups were using commercial CCD controllers which were exchanged for low-noise custom-made readout electronics developed for DAMIC-M and briefly mentioned in Section\,\ref{sec:electronics}.

\section{Detector Description}
\label{sec:design}
The LBC is designed to provide $O(\mathrm{1})$\,dru background environment for the operation of CCDs with modular electronics. The innermost region of the detector consists of a cold copper housing, which simultaneously serves to cool the CCDs and provide an infrared shield from the surrounding room temperature materials. This cold copper housing is enclosed by 2\,cm of ancient Roman lead from a sunken galley\,\cite{Lead_at_LSM,EDELWEISS:2013wrh}, then an additional 4-10\,cm of very low background ("tr\`{e}s faible activit\'{e}" or TFA) lead. Kinked channels provide a pathway for the two-layer polyimide flex cables, used for CCD operation and readout, to exit the inner region eliminating a line-of-sight to the inner copper housing. Cold OFHC\footnote{In reality, it is from OFE-OK copper grade by Luvata in Finland. Because it is comparable to oxygen-free high thermal conductivity copper used in other experiments, we will use acronym OFHC.} copper "fingers" travel through four narrow gaps in the interior (warm) lead shielding with $\sim$1\,mm clearance to provide thermal contact and mechanical support for the inner copper housing.

Outside of the interior lead shielding are four axially-spaced intermediate boards, which provide the first stage amplification to the CCD video signal. The boards connect to flex cables through commercial 50-pin connectors by TE Connectivity on each side. The output side of each board connects to a 45\,cm long, five-layer CCD flex cable that exits below the lead shielding to commercial 50-pin D-Sub vacuum feedthroughs by Accu-Glass below.

The cryostat is designed in two pieces, each of them machined from OFHC copper. A top can with an inner diameter of 40\,cm and height of 25\,cm can be lifted up and away from the detector with a crane for easy access to the CCDs. A bottom cross provides structural support as well as all vacuum and electronics feedthroughs. A set of lead discs are affixed to the top can to shield the CCDs from above, and are lifted during opening.

\begin{figure*}[!t]
    \centering
    \includegraphics[width=0.9\textwidth, trim=0 0 0 0, clip=true]{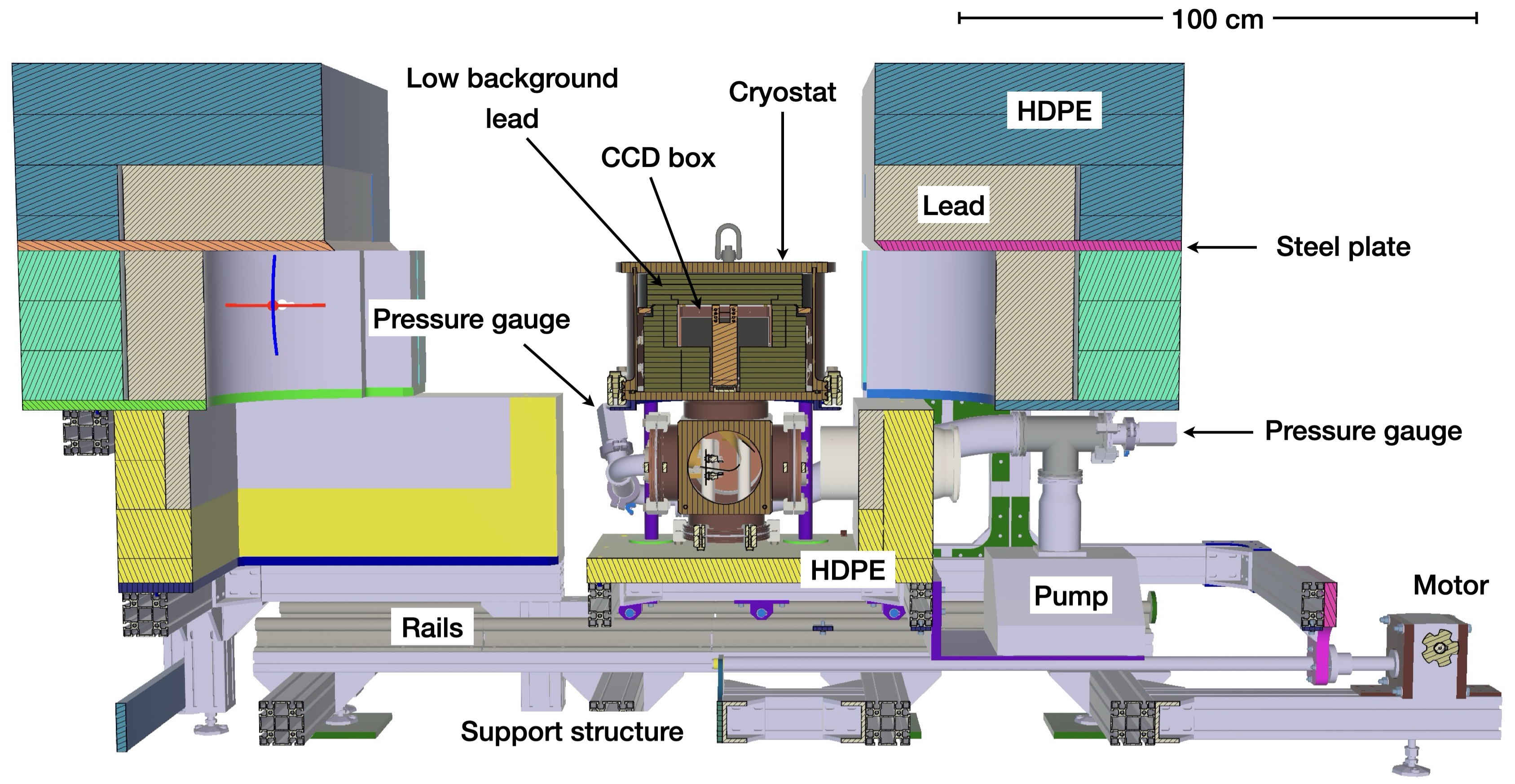}
    \caption{A cross-sectional view of the LBC detector 
    in its fully opened position with major parts labelled. The cryostat and right side of the external shield can be moved on rails, while the left side of the external shield is static. Other parts of the detector are labelled.}
    \label{fig:LBC}
\end{figure*}

The cryostat is housed inside of a custom shielding support structure built out of aluminum extrusions and steel plates. This structure includes a mechanism to horizontally open the shielding by sliding half of the shielding and the cryostat away from a fixed shielding wall. Outside of the cryostat, the cylindrical shielding consists of lead and high-density polyethylene (HDPE). Fig.\,\ref{fig:LBC} shows a cross sectional cut through the CAD model of the LBC in the open configuration.

The design allows for easy exchange of CCDs and internal electronics, as well as for the deployment of calibration sources. This provides an efficient way to characterize various CCDs and front-end electronics in different background environments.

\subsection{Charge-Coupled Devices}
In the originally commissioned configuration, called Setup\,1, two CCDs -- each of the pixel format 6144 columns and 4128 rows or 6k$\times$4k -- were installed in the LBC in late 2021\footnote{A couple of CCDs of the same format were also installed in the upgrade of the DAMIC detector at SNOLAB\,\cite{DAMIC:2023ela}.}. These CCDs feature a three-phase polysilicon gate structure with a buried p-channel, with a pixel size of 15$\times$15\,$\mu$m$^2$. The CCD thickness after all fabrication steps is 669\,$\mu$m\,\cite{DAMIC:2021crr}, totaling $\sim$8.9\,g in the active mass. The bulk of these device consist of high-resistivity ($>$10\,k$\Omega$\,cm) n-type silicon
which allows for fully depleted operation at substrate bias voltage V$_\mathrm{sub}\geq$40\,V. The substrate voltage is applied to the backside of the silicon substrate and ensures that the entire thickness of the substrate is free of free charge carriers\,\cite{Holland:2009zz}.

The devices were developed by Lawrence Berkeley National Laboratory (LBNL)\,\cite{Holland:2002zz,Holland:2003zz,Holland:2009zz} and fabricated by Teledyne DALSA Semiconductor with front-side polishing at Omnisil. The 150\,mm diameter silicon wafers were sliced from an ingot pulled by Topsil GlobalWafers\ A/S. Wirebonding and packaging were done at the University of Washington.

These 6k$\times$4k CCD have two skipper amplifiers on one side and two standard amplifiers on the opposite side. Unlike conventional CCD amplifiers, skipper amplifiers can be configured to make multiple, non-destructive charge measurements (NDCMs)\,\cite{skipper,Chandler1990zz,Tiffenberg:2017aac,DAMIC-M:2022xtp}. Skipper readout essentially moves the charge contained in each pixel back-and-forth into the readout node allowing for many measurements of the same pixel so that they can be averaged. Since the measurements are uncorrelated, the readout noise is reduced to $\sigma_1/\sqrt{N_{\rm{skip}}}$, where $\sigma_1$ is the single-sample readout noise (the standard deviation of a single charge measurement) and $N_{\rm{skip}}$ is the number of NDCMs. By taking a large enough number of NDCMs, the readout noise can reach the sub-electron level and single charges\footnote{For the sake of clarity, we use the term electrons to indicate charge carriers detected in the CCD. However, holes are held in the pixels of p-channel LBNL CCDs.} are resolved. The single-electron resolution also provides a straightforward way of calibrating the energy response of the detectors\,\cite{DAMIC-M:2022xtp}.

A two-layer CCD flex cable carries the clock voltages to and video outputs from each CCD. The cable is glued to an extended non-functional area of the CCD. To improve adhesion, rectangular silicon shims were glued between the CCD die and the flex. The glue is low-background Epotek 301-2 previously selected for DAMIC at SNOLAB.
The cable lines are wedge bonded with 25-$\mu$m thick Al wire to corresponding pads on the CCD. Special electrostatic discharge (ESD) safety precautions, described in Sec.\,\ref{sec:design_infrastructure}, have been implemented during handling, packaging and testing CCDs to reach high packaging yield. During this time, care is taken to minimize the exposure of the CCDs to the air and hence, the accumulation of radon progeny on the CCD surface.

CCDs were characterized in a surface laboratory and only the devices passing our science-grade criteria were selected for the LBC. The selection criteria are as follows: all amplifiers work and have low readout noise; none or only few defect pixels are in the pixel area; minimal clock-transfer inefficiency during vertical and horizontal clocking of charge; single electron resolution by skipper amplifiers; and the CCD has low leakage current. In addition, dedicated measurements with atmospheric muon tracks were taken to model the diffusion of charge as it drifts from the back or the bulk to the buried channel on the front side of the CCD\,\cite{PhysRevD.94.082006}.

\begin{figure*}[!t]
    \centering
    \includegraphics[height=0.43\textwidth, trim=0 0 0 0, clip=true]{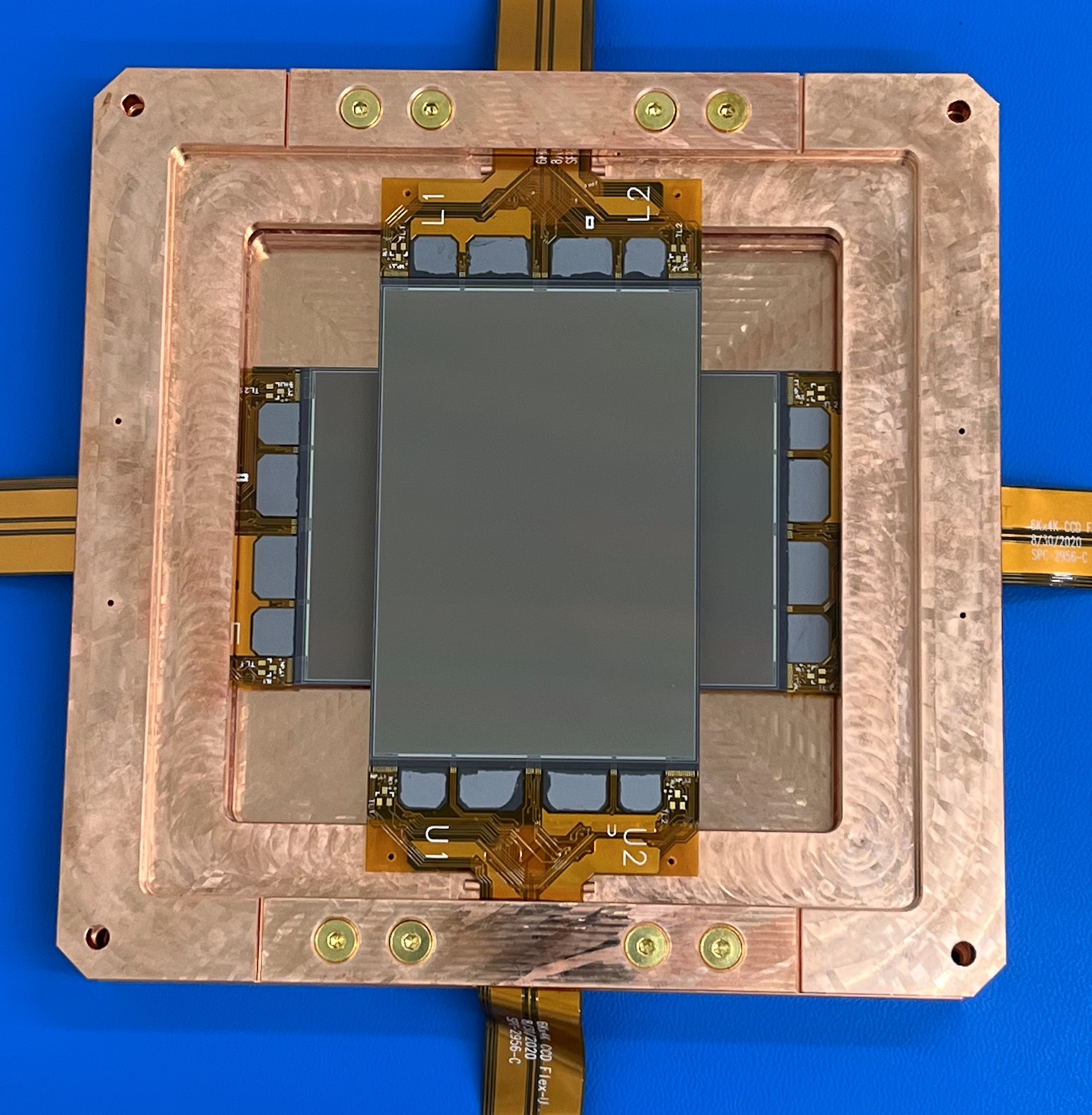}
    \includegraphics[height=0.43\textwidth, trim=0 0 0 0, clip=true]{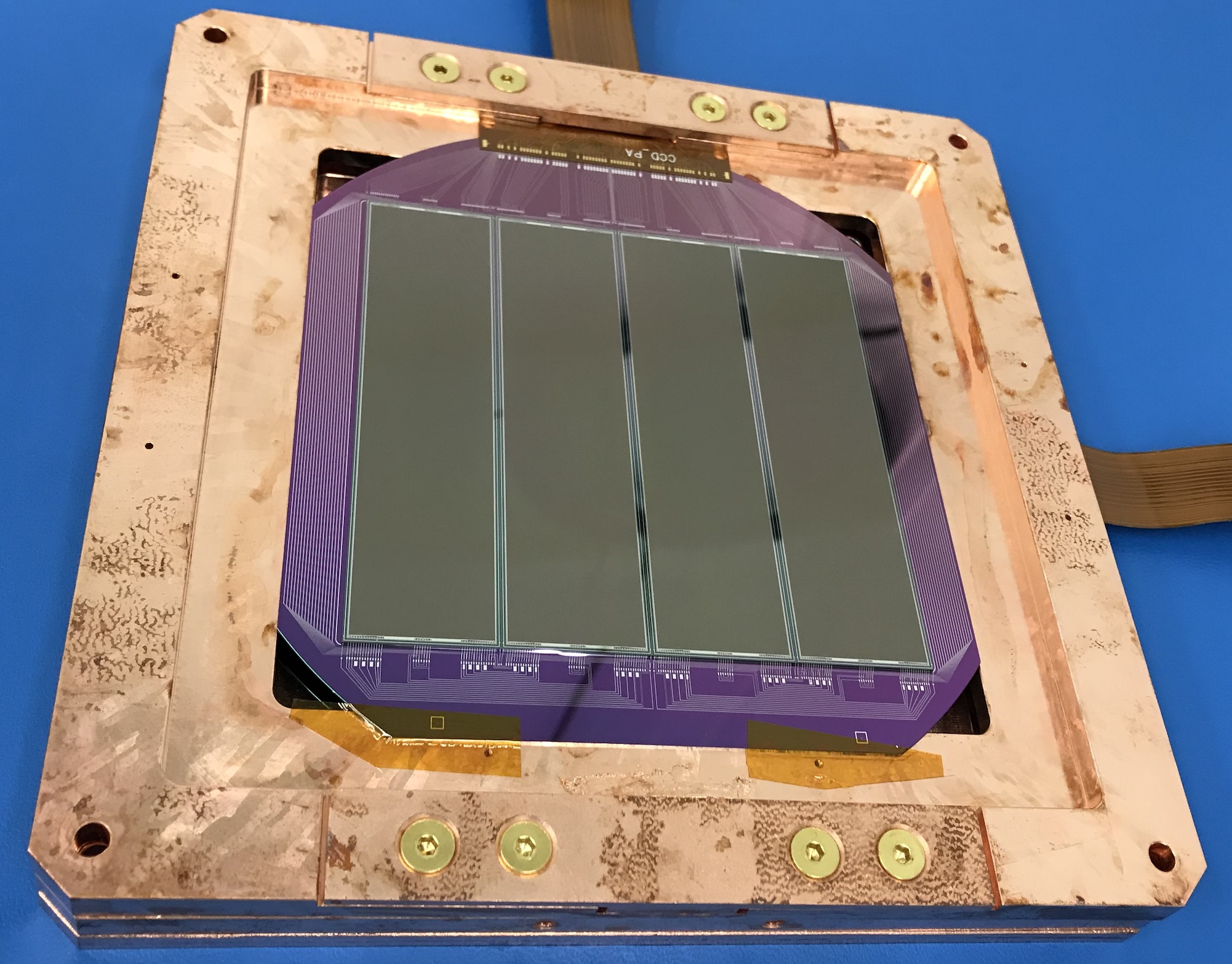}
    \caption{(Left) Two 6k$\times$4k CCDs installed in the OFHC copper CCD box. Note, that there is no material between the bottom and top CCD. Two flex cables from each CCD are fixed with copper t-bars on the way out of the box. (Right) Two CCD modules installed in the same CCD box as in (left), but only the the top module can be seen. 6k$\times$1.5k CCDs are glued on a silicon pitch adapter with traces bringing voltages to all four CCDs from the cable. Note that the flex from the second module is visible in the right-hand side photo.}
    \label{fig:CCDs}
\end{figure*}

In the LBC, the two CCDs were placed in a 180$\times$180\,mm$^2$ interlocking copper box.
This CCD box consists of one frame for each CCD, top and bottom lids and t-bars (to clamp the flex cable to the frame). The parts are tightly connected with brass screws to prevent infrared radiation and to press CCD cables coming out of the CCD box with t-bars. Fig.\,\ref{fig:CCDs} (left) shows the CCD configuration before installing the top lid.

In February 2023, two DAMIC-M CCD prototype modules were installed for characterization in a low background environment. Each module includes four CCDs with an active pixel area of 6144 columns and 1536 rows (6k$\times$1.5k) glued to a silicon pitch adapter with traces for clocks, biases and video signals. The pitch adapters were fabricated at the Washington Nanofabrication Facility (WNF) from high-resistivity wafers with 0.5\,$\mu$m of thermal oxide. The traces are made from 1-$\mu$m thick aluminum and their width/spacing were optimized to minimize electrical resistance and parasitic capacitances.
In the CCD module, only one amplifier is read out from each CCD. The module installed in the copper frame can be seen in Fig.~\ref{fig:CCDs} (right). By installing these two CCD modules in the LBC, we were able to verify the design and characterize 6k$\times$1.5k CCDs and their electronics in a low background environment.

\subsection{Boards and Cabling}
Here, we will describe the electronics chain used for the 6k$\times$4k CCDs originally installed in the LBC and provide a few details about the configuration used for the CCD modules. All components were tested in laboratories at various institutions before their installation at LSM.

The electronics chain starts at the two-layer CCD flex cable, which is wire bonded directly to two sides of the 6k$\times$4k CCD. This type of a flex cable has been developed to achieve low background in future experiments, like DAMIC-M\,\cite{Arnquist:2023gtq}. The flex cables used for the LBC have the same layers but they are not as radiopure as those described in\,\cite{Arnquist:2023gtq}, because they were fabricated without costly cleaning steps.

Each 17\,cm long CCD cable is routed through a thermally insulating polychlorotrifluoroethylene (PCTFE) sleeve to the outside of the inner lead shielding castle. A 50-pin D-Sub connector at the end of the cable is plugged into a intermediate PCB. As these boards and D-Sub-50 connectors are not radiopure, they are shielded from CCDs by the internal lead.

The intermediate board has an amplifier, bias line filters, clock shaping circuit and two D-Sub connectors. Right angle connectors were used at the beginning, but boards with straight connectors were installed later after comparing radioassay results for both type of connectors.

A second-stage 45\,cm long flexible cable, still thermally isolated from warm lead, goes from the intermediate board down to the cross part of the cryostat, where it is plugged into a two-to-one board. This board merges two cables coming from opposite sides of the same CCD and connects to the vacuum feedthrough. In early 2023, when two CCD modules replaced the 6k$\times$4k CCDs, new intermediate boards were installed. Because each CCD module has only one cable to carry all signals, previously used two-to-one boards were taken out.

\subsection{Cold Copper Parts and IR Shielding}
\label{sec:cryostat}

The CCDs are operated at low temperature to achieve low dark current and in vacuum to avoid humidity condensation, which may cause shorts between exposed conductive parts. Typically, the LBC operates the CCDs at $\sim$130\,K. Images have also been acquired at higher temperatures (up to 180\,K) in runs dedicated to the identification of hot pixels and columns and studies of temperature dependence of the dark current. Low, stable temperatures are necessary to optimally operate the CCDs, thus imposing strict requirements on cold copper parts.

The CCDs are housed in a copper box which cools the devices and acts as an IR shield. The dominant thermal path to the CCDs is through the CCD flex cables which are clamped to the CCD box via t-bars. In this configuration, the temperature remains stable when the on-chip CCD amplifiers are powered as detailed in Sec.\,\ref{sec:pressure_temperature}.

\begin{figure*}[!t]
    \centering
    \includegraphics[width=0.8\textwidth, trim=0 0 0 0, clip=true]{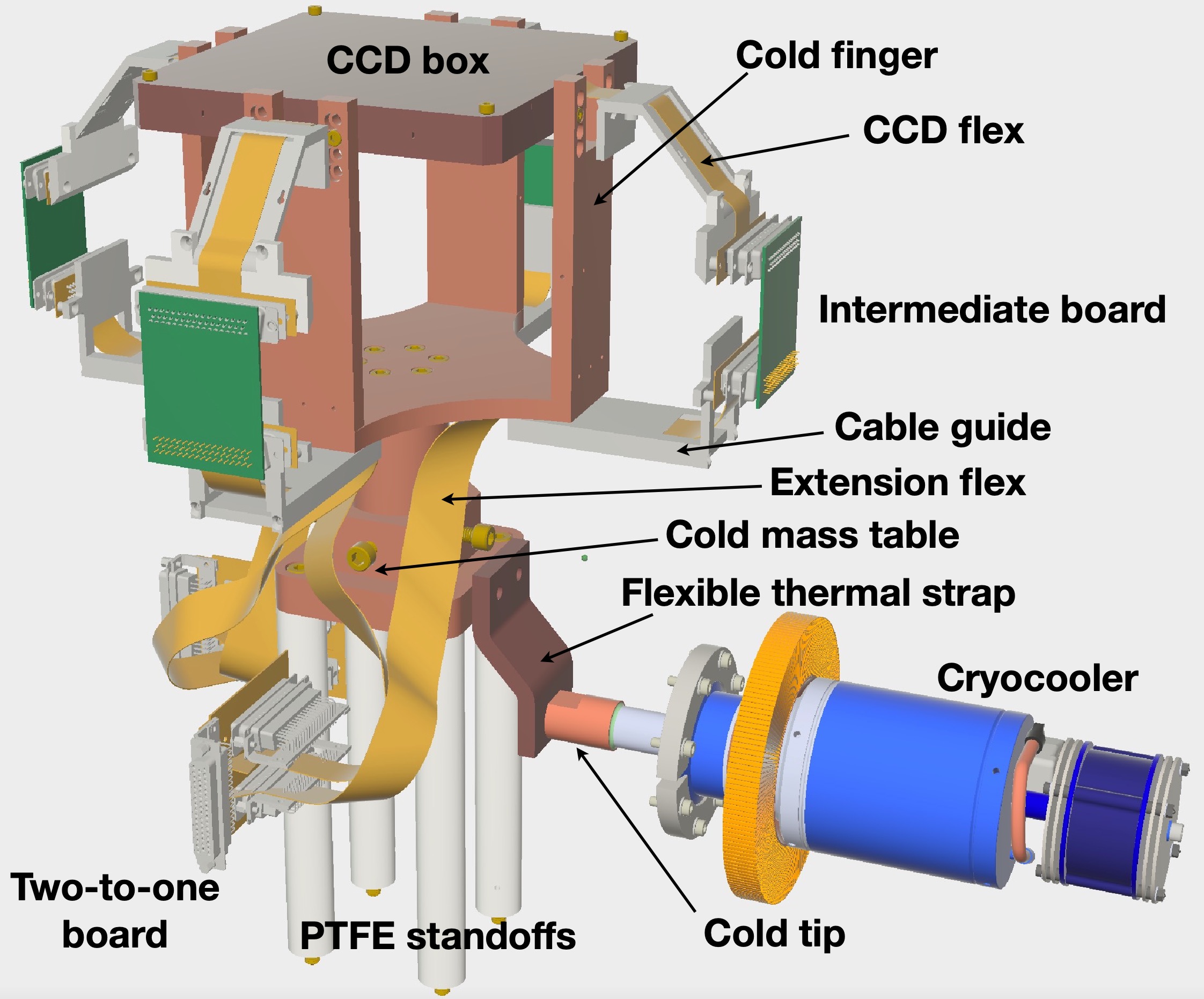}
    \caption{All parts inside the cryostat except the inner lead shielding, RTDs and heaters; together with the cryocooler (without its air shroud). The CCD box is lifted above other components to have enough lead below it. Cold copper pieces connect to the cold tip of the cryocooler (right, bottom). Four extension flex cables encapsulated in thermal shielding can be seen running from the CCD box to intermediate boards. The second stage cables run down from the boards to a couple of two-to-one boards which are plugged to the vacuum feedthrough. Lead, temperature sensors, and heaters are not shown in the figure for clarity.}
    \label{fig:ColdCopper_Cabling}
\end{figure*}

The CCD box is supported by, and thermally coupled to, four cold fingers which rise from a copper hand. This hand is atop of an arm which is fixed to a cold mass table. The cold mass table sits on four polytetrafluoroethylene (PTFE) standoffs, which are connected to the bottom flange with vented studs. The drawing of the setup can be seen in Fig.\,\ref{fig:ColdCopper_Cabling}. All these copper parts are kept at cold during the operation and they have sufficient clearance to other parts inside the cryostat that are kept at room temperature.

The connection to the cold tip of a cryocooler is made by a commercial flexible thermal strap by Thermal Management Technologies. Individual copper pieces were tightly connected with brass screws and a thin layer of thermally conductive grease (i.e. Apiezon\ N cryogenic high vacuum grease) was applied between them. For an electrical isolation, two 25\,$\mu$m-thick Kapton tapes are placed between the cold tip and flexible strap.

The copper parts inside the cryostat are made from OFHC copper from Luvata and were machined at the University of Z\"{u}rich. The cryostat was made by CINEL in Italy from the same type of copper. The cold copper parts and electronics chain inside the cryostat, support structure, and a connection to the cryocooler are shown in Fig.\,\ref{fig:ColdCopper_Cabling}. For the clarity, we do not show the internal lead shielding.

Because the copper lids are directly above and below CCDs, they are a major source of radioactive background. OFHC copper lids were used while taking science data with Setup\,1 in 2022\,\cite{DAMIC-M:2023gxo,DAMIC-M:2023hgj}, but were taken out later the same year. The new installed lids are from less radioactive, electro-formed copper (EFC) grown and machined by Laboratorio Subterraneo de Canfranc (LSC)\,\cite{Borjabad:2018wda}. The EFC lids not only have much lower radioactive contamination than the commercial OFHC copper, but also their cosmogenic activation time has been reduced from many months to less than a week as described in Sec.\,\ref{sec:activation}.

\subsection{Internal Passive Shielding}
A lead castle inside the cryostat provides shielding against high energy gamma radiation originating from electronics or any other part inside or outside the cryostat. The castle is built in such a way that there is at least 6\,cm of internal lead shielding the CCD box from the bottom and sides. The innermost 2\,cm is made from ancient Roman lead provided by LSM\,\cite{Lead_at_LSM,EDELWEISS:2013wrh} and the rest is from TFA lead casted by Fonderie de Gentilly in France.

Three lead disks, each 2\,cm thick, are attached with brass studs and a copper ring to the inside of the cryostat top cylinder above the CCD box. The bottom one directly above the CCD box is from ancient Roman lead. Six, 2\,cm-thick square plates are below the CCD box with the top one from ancient lead. Two of these plates can be removed without affecting the background level if an upgrade to a second CCD box will be needed. Vertical pieces on all four sides are designed to accommodate the thermal guides for CCD cables while providing sufficient shielding.

\begin{figure*}[!t]
    \centering
    \includegraphics[width=0.8\textwidth, trim=0 0 0 0, clip=true]{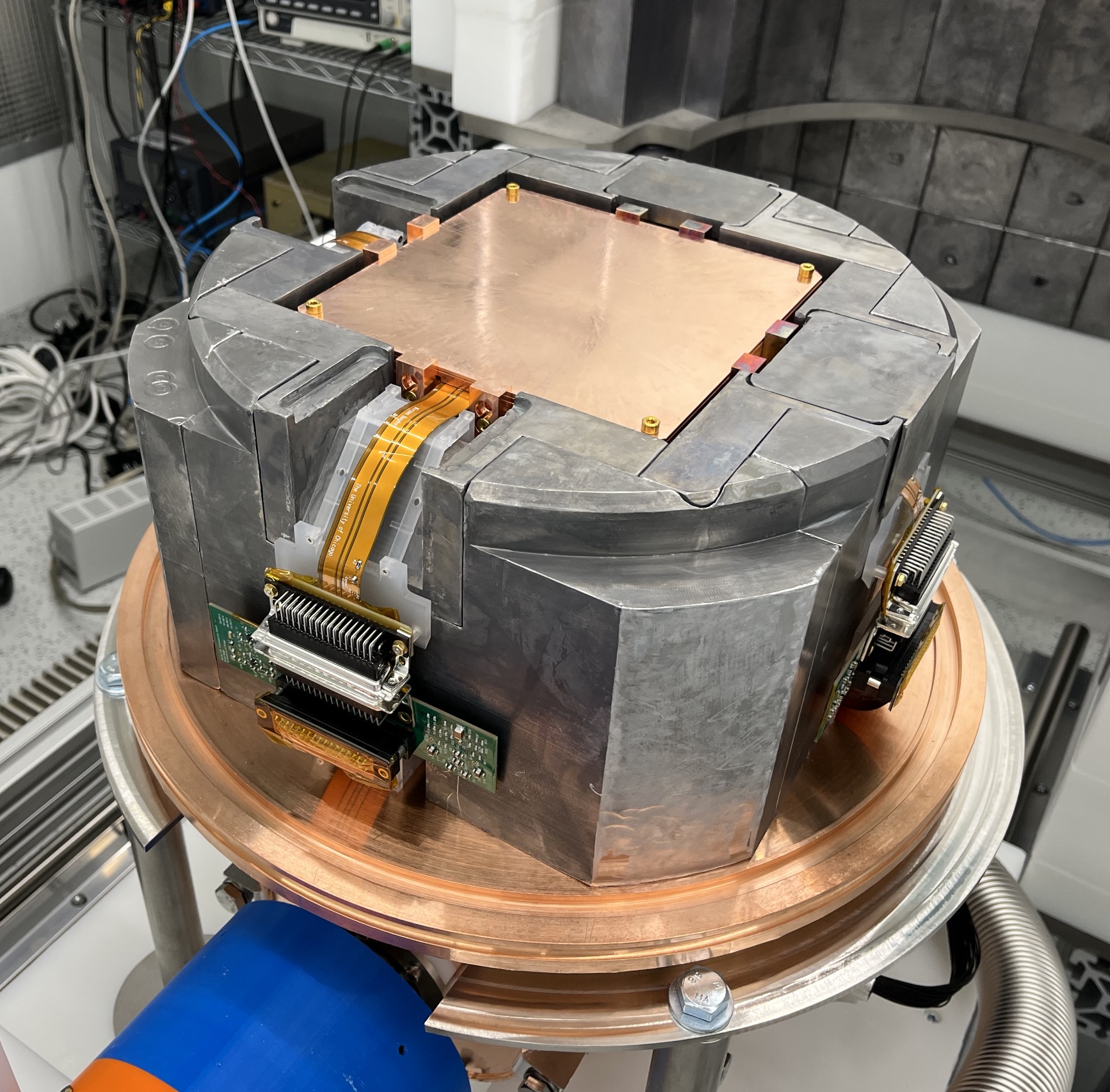}
    \caption{The CCD box inside the lead castle. The CCD cable runs in a PCTFE cable guide to keep it thermally isolated from lead and is plugged to the intermediate board. In this photo, the intermediate boards have right angle connectors, but new boards with straight connectors were installed later.}
    \label{fig:InnerLead}
\end{figure*}

The lead castle is kept at room temperature because it is thermally isolated from cold copper parts. This was achieved either by using materials with low thermal conductivity or keeping a few millimeters of clearance between lead and cold copper parts.
The internal lead shield is shown in Fig.\,\ref{fig:InnerLead}.

\subsection{External Passive Shielding and Support Structure}

\begin{figure*}[!t]
    \centering
    \includegraphics[width=0.8\textwidth, angle=0, trim=0 0 0 0, clip=true]{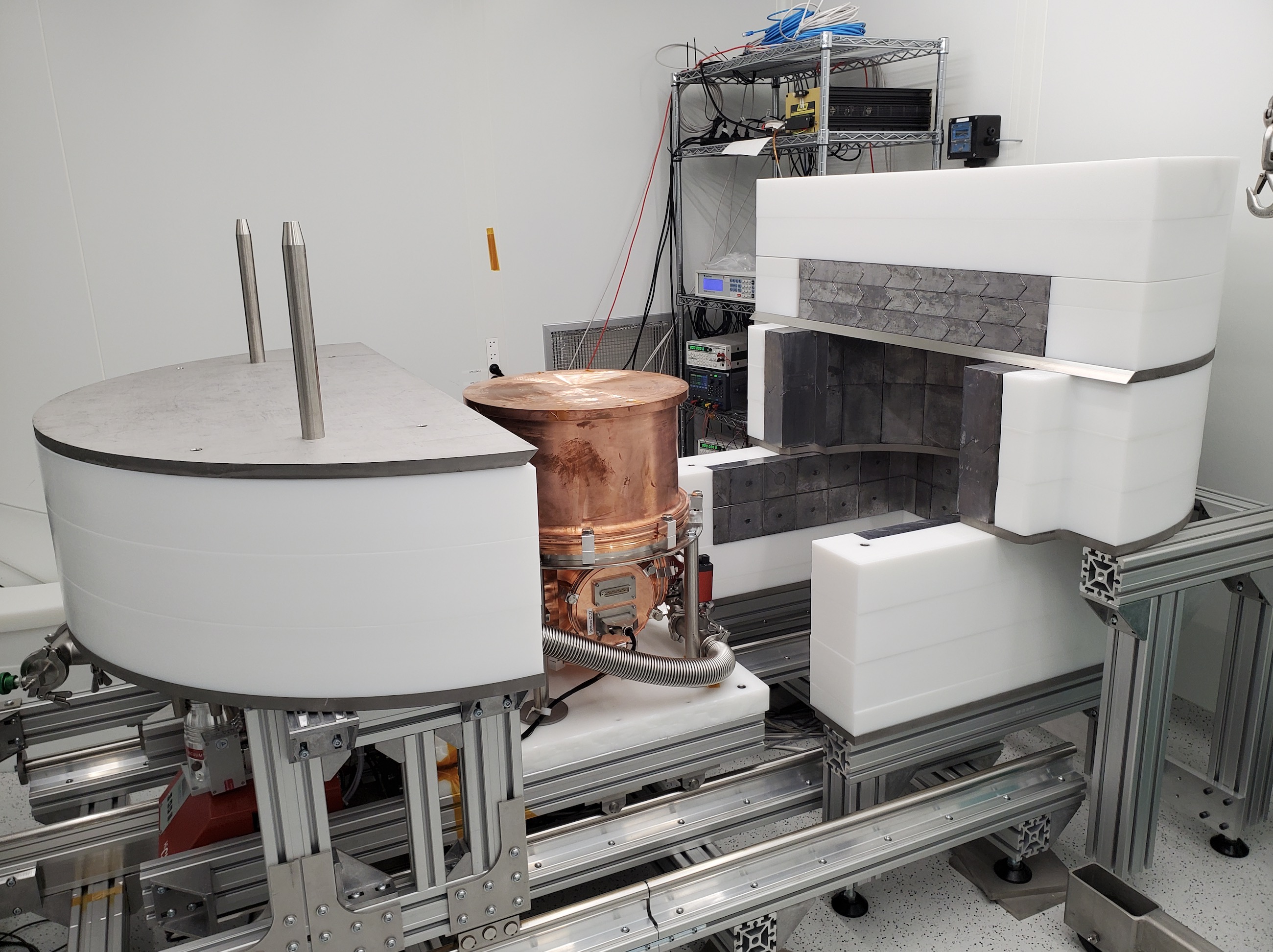}
    \caption{The LBC is shown in its open configuration with partly disassembled external shielding. The cryostat in the middle is surrounded by lead and HDPE. The static part of the support structure is on the right and the left side moves on rails. One feedthrough flange and parts of the vacuum system are also visible.}
    \label{fig:DetectorWhole}
\end{figure*}

External gamma radiation and neutrons are attenuated by a layer of low background lead (faible activit\`{e} or FA) from LEMER and HDPE by Abaqueplast, both in France. The shielding around and above the cryostat cylinder, where the shielding of CCDs is the most important, consists of 15 and 20\,cm of lead and HDPE, respectively, see Fig.\,\ref{fig:LBC} and \ref{fig:DetectorWhole}. The bottom cross of the cryostat is surrounded by 5\,cm of lead from all sides but below the cryostat and 10\,cm of HDPE. The 41\,cm diameter ring of lead shielding above the cryostat together with HDPE half-disks is supported by a 2\,cm thick stainless steel plate.

The support structure standing on the cleanroom floor is static while a motorised opening system can move half of the shield on rails, allowing easy access to the cryostat. Neither the motor nor rails use lubrication and thus do not compromise the cleanroom cleanliness. A photo of the fully assembled and partly opened detector is in Fig.\,\ref{fig:DetectorWhole}.

A movable crane is kept inside the cleanroom. It was employed during the assembly of the support structure,  external shielding and cryostat. It is utilized to lift the cylindrical cover any time access inside the cryostat is required.

\subsection{LSM Infrastructure}
\label{sec:design_infrastructure}
A new cleanroom was built for DAMIC-M on the ground floor of LSM between the neutrinoless double-beta decay experiment SuperNEMO\,\cite{SuperNEMO:2010wnd} and a room with high purity Germanium (HPGe) detectors\,\cite{Loaiza:2011zza}. In the future, the LBC will be removed so that DAMIC-M can be assembled in this cleanroom.
The total floor area is 6.2$\times$4.3\,m$^2$. The main room is class ISO\,5 for air cleanliness and the gowning room is ISO\,6. The room has a roof opening to load big and heavy pieces with the LSM overhead crane. This option was used before the assembly of the LBC and the room was cleaned afterwards.

The furniture and equipment inside the cleanroom are ESD safe to comply with requirements on handling ESD-sensitive equipment, like CCDs and electronics.
While working with CCDs, we require the humidity above 35\% and we deploy a local humidifier using deionized water if needed.

The radon concentration has been continuously measured every 10 minutes with Durridge RAD7 in the cleanroom since January 2023. The median radon concentration is (18.4$\pm$8.5)\,Bq/m$^3$ with 99\% of the data below 42\,Bq/m$^3$. Other backgrounds were measured by different groups as part of the characterisation of the radiation environment in the laboratory, and they are: gamma flux of 3.8$\times$10$^{-6}$\,cm$^{-2}$s$^{-1}$ between 4 and 6\,MeV\,\cite{NEMO:2002zps}, thermal neutron flux of $4.6\times10^{-6}$\,n/cm$^2$/s at the LBC location\,\cite{Rozov:2010bk}, and fast neutron flux above 1 MeV of (1.06$\pm$0.60)$\times$10$^{-6}$\,n/cm$^2$/s\,\cite{EDELWEISS:2006uee}. Details on these measurements can be found in given references.

\section{Detector Readout and Instrumentation}
\label{sec:instruments}

The LBC is remotely controlled and monitored by custom data acquisition systems placed outside the external shield. The LBC is remotely controlled and monitored by custom data acquisition systems placed outside the external shield. We describe two different electronics systems installed in the LBC to run CCDs, of which the first one collected the data published in two DAMIC-M publications\,\cite{DAMIC-M:2023gxo,DAMIC-M:2023hgj}. The slow control, DAQ software and data quality monitor system are also detailed.


\subsection{Readout Electronics}
\label{sec:electronics}

A custom data acquisition system based on a commercial Gen-III CCD controller from Astronomical Research Cameras, Inc. (ARC) was used for initial tests with the LBC and to collect data used in two papers\,\cite{DAMIC-M:2023gxo,DAMIC-M:2023hgj}. The controller provides the voltage biases and clocks required for operation of the CCD and reads out the video signal from two CCD amplifiers. External low-noise power supplies provide DC voltages for controllers, amplifiers and V$_\mathrm{sub}$. Two CCD controllers were used to read two 6k$\times$4k CCDs or two CCDs in two CCD modules.

The signals for the CCDs are generated by a clock (ARC-32) and bias (ARC-33) board and the CCD signal is read by a video board (ARC-45). Each board has a separate cable, which are merged by a custom-made breakout board to a 50-pin connector. The ARC controller communicates to the computer via an ARC-22 board in the controller crate and an ARC-66x in the PC. A custom-made, 3\,m-long cable harness of 50\,$\Omega$ micro-coaxial cables (model 40322-001 by Hitachi) brings signals to the vacuum feedthrough in the cryostat.

All components of the electronics chain are electrically grounded and kept isolated from other devices to minimize the noise level on the readout signal and cross talk.

The new DAMIC-M readout electronics were successfully installed in the LBC in January 2024 and are used to control CCDs since then. These electronics were developed for DAMIC-M, have lower readout noise than the ARC CCD controller and can read all four CCDs in a module. The electronics chain consists of a custom-made Acquisition and Control Module (ACM) plugged to a VME crate (Wiener VME 6023x610), a 3\,m long cable by Samtec and a custom-made front-end board. In addition, an adapter board connects the 3\,m cable and D-Sub-50 connector in the vacuum feedthrough. More details on the new electronics will be given in a separate publication.

\subsection{Instrumentation}
\label{sec:instrumentation}
As described in Section\,\ref{sec:cryostat}, the CCDs are housed inside a light-tight copper cryostat. The nominal operating temperature is approximately 130\,K, providing a low dark current and high charge carrier mobility. This is achieved by using an air-cooled Sunpower CryoTel GT cryocooler. It is equipped with an active balancer to minimize mechanical vibrations and prevent microphonics noise. Airflow for the cooling of the cryocooler is provided by a 3D-printed tube and a PC fan. The cryocooler assembly is mounted to a dedicated flange on the cryostat. The cryocooler's cold tip is connected to the cold copper (see Sec.\,\ref{sec:cryostat}), maintaining electrical isolation via Kapton tape.

The temperature of the cold copper is monitored by resistance temperature detectors (RTDs) PT-103 epoxied in cold fingers and cold mass table, see Fig.\,\ref{fig:ColdCopper_Cabling}. The cooling rate and temperature are set by a Lakeshore\,336 controller that monitors the RTDs and controls the power to a 100\,W heater installed 2.5\,cm next to the RTD closest to the cryocooler.
To minimize temperature-induced stress on CCDs, we use the cooling rate of 0.1\,K/min during temperature cycles. The system maintains the temperature within $\pm$0.2\,K for many months, see Sec.\,\ref{sec:ccd_performance}.

The cryostat is kept at a pressure of $\sim$5$\times$10$^{-6}$\,mbar with a Pfeiffer HiCube 80 turbo pump. Two Pfeiffer PTR 91 pressure gauges, which are controlled by a Pfeiffer TPG362 controller, are used to monitor the system pressure in different locations. One gauge is connected to the vacuum tube next to the pump and the other next to the vacuum connection to the cryostat. Both pressure gauges are turned off during data taking because we have observed an increase of the dark current in the CCDs when they were turned on.

The components of the electronics chain (see Sec.\,\ref{sec:electronics}) are powered by independent, low-noise power supplies. Specifically, 3-channel Keysight E3612A's are used to supply power to the CCD controller bias boards and the second-stage amplifiers. A high voltage SRS DC205 provides V$_\mathrm{sub}$ to the CCDs.

The detector and its components are protected by two-levels of backup power: (1) the DAMIC-M APC 2200 UPS and (2) the LSM system-wide UPS. In the case of a power outage, both UPS systems would be able to operate from battery power for at least one hour and supply instruments as shown in Fig.\,\ref{fig:SC_schematic}. This is adequate time for remote shifters to respond to alarms (see Sec\,\ref{sec:sc}) and safely shut down the experiment. In the unlikely scenario of an outage longer than one hour with no intervention from shift scientists, instruments would undergo a hard shut down. We have verified that the transition to UPS power is smooth and maintains the SC and DAQ systems up in the case of a power cut.

\subsection{Slow Control System}
\label{sec:sc}
All of the instrumentation discussed in Section\,\ref{sec:instrumentation} required to run the cryostat and electronics are controlled and monitored by a custom slow control system. The system was adapted from astro-slow-control\,\cite{Nikkel_ASC} developed for the LUX experiment, and centers around a MySQL database populated by independent processes for telemetry. Each instrument has a process that stores input/output data to/from the device. Additional processes are used as watchdog daemons, alarm systems, and other safety features. The backend C programs encode the different communication protocols required to transmit and receive messages from the various instruments. A PHP web frontend interface is used for real-time monitoring, low-level analysis, and setting new control parameters to the backend programs.

\begin{figure*}[!t]
    \centering
    \includegraphics[width=0.99\textwidth, trim=0 0 0 0, clip=true]{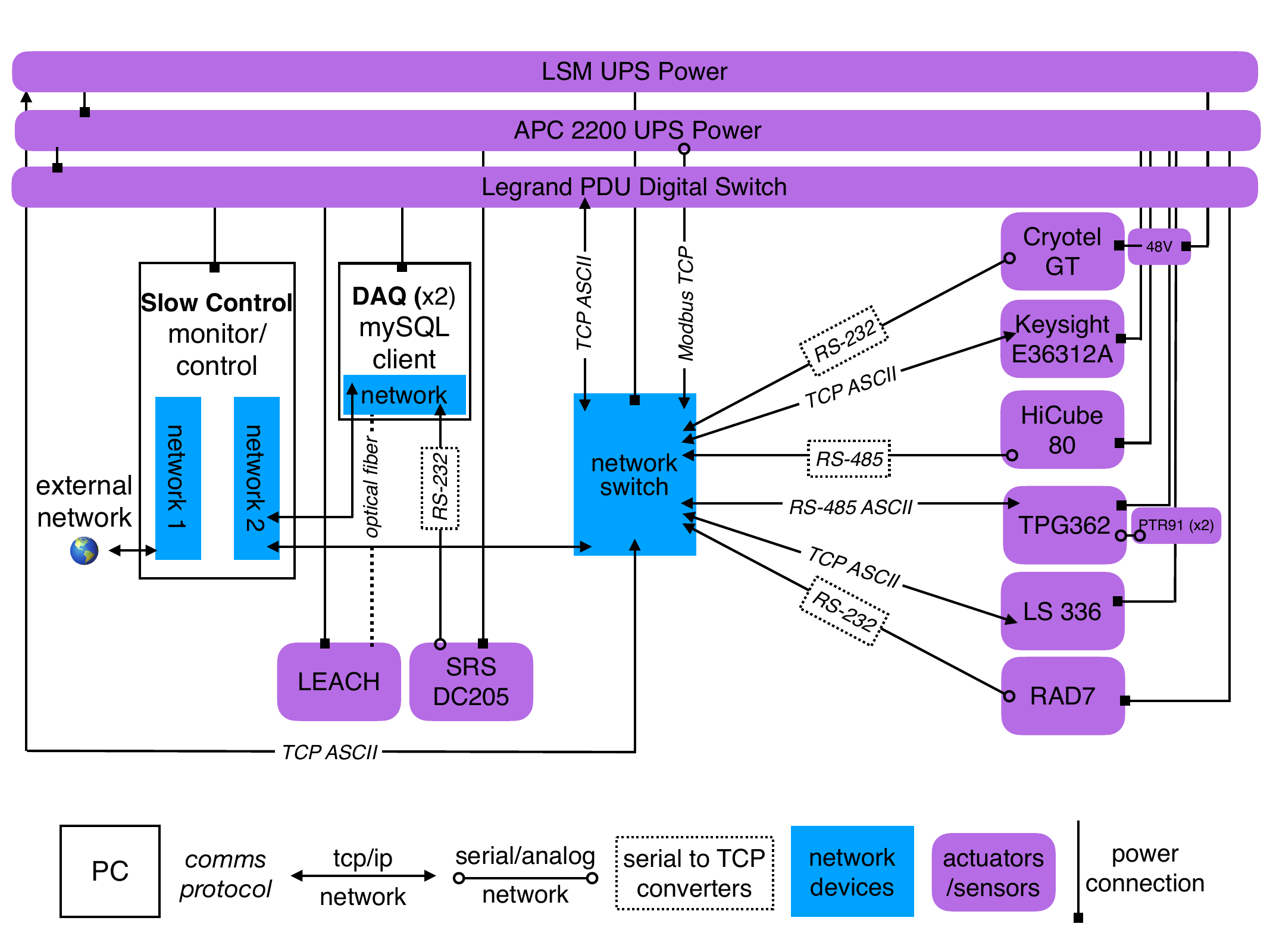}
    \caption{Schematic of the slow control system. This digital system is controlled by a Slow Control PC through a series of TCP networks (blue) to the auxiliary devices (purple). The backend C programs control I/O from a MySQL database. The system was made to be user-friendly through a PHP web interface. This is for the original setup with ARC CCD Controller (LEACH). LS\,336 stands for Lakeshore temperature controller.}.
    \label{fig:SC_schematic}
\end{figure*}

Communication between the slow control system and the instrumentation is based on a private internal network, as illustrated in Figure\,\ref{fig:SC_schematic}. A host PC running CentOS8 connects to both the external world for web access and the private TCP/IP slow control network. The network then plugs into a large Ethernet switch that interfaces with all of the instruments. Some of the devices can directly communicate through this line via TCP/IP protocols if they have physical Ethernet ports, while others must be converted via serial device servers to translate other protocols, e.g. RS-232. The DAQ PC (Sec.\,\ref{sec:daq}) has the ability to interface with the slow control system via a MySQL client.

Conditions that can pose a hazard for the operation of the CCDs are triggered on by alarm daemons. For example, a failure of the vacuum pump would cause moisture to condense on the cold CCDs and introduce shorts in the electrical components. To prevent such a scenario, the pump is powered by the local UPS which is monitored by the slow control system as is shown in Fig.\,\ref{fig:SC_schematic}. These alarms signal to on-shift users that the CCDs must be promptly shut down. In addition, a safety watchdog daemon is used to execute a shutdown if any pump parameters are beyond user-defined thresholds for more than two consecutive measurements (occurring every 5\,s). Additionally, sound and light alarms would be triggered and notifications sent to system managers.

\subsection{DAQ software}
\label{sec:daq}
Each ARC controller communicates via an optical fiber to the DAQ computer through the custom-built CCDDrone\,\cite{ccddrone} software, which sends configuration files, executes the readout sequence and receives the output images. Both are controlled centrally by the ccd-cdaq\,\cite{ccd-cdaq} Python library, designed to control CCD experiments. It loads another Python library, libabcd\,\cite{libabcd}, which implements the MQTT protocol to communicate between processes and logs it. For debugging purposes, all messages that pass through the MQTT broker. A set of four applications (run control, DAQ status monitor, CCD manager, and spy) that run continuously in the background send and receive messages in an orderly fashion. Two CCD client applications, one for each CCD controller, communicate with each CCDDrone program using a custom Python library.

The operator sends commands directly to the central DAQ, which automatically adds the commands to a queue. Each command is executed sequentially, with the next command waiting for the previous one to return. The operator can pause or clear the queue at any time, as well as send a series of commands via a script. They can also decide whether each command applies to a specific CCD client or to all, and the central DAQ sends the messages to the appropriate client. This makes it possible to configure each of the CCDs in a different way and then start the readout of both of them at the same time. 

As shown in Fig.\,\ref{fig:SC_schematic}, the DAQ machine runs the CCD control software. It features two network cards for internet connection and TCP communication with the power supply that provides V$_\mathrm{sub}$. The latter allows to perform all operations related with V$_\mathrm{sub}$ in an automated manner.

The described DAQ software is used also for new ACM controllers installed in the LBC recently, see Sec.\,\ref{sec:electronics}. To achieve this, we have written new firmware and modified local DAQ software.

\subsection{Data quality monitor}
An automated data quality monitor (DQM) system has been developed for the LBC to ensure detector functionality and operational efficiency. The DQM performs periodic data quality checks and spots acquisition problems and/or malfunction, whether in the detector hardware or reconstruction software, and issues corresponding alarms. This proactive approach enables rapid intervention to minimize downtime and maintain uninterrupted data collection.

The DQM framework provides tools to generate, store and visualize histograms and scalar elements (monitoring elements), using MongoDB as the database backend. The first step is low-level data processing of raw images to obtain the charge value for pixels, calculate median baselines, resolution, dark current per row, etc. The time evolution of these parameters is monitored. For example, the stability of the dark current can be used to identify new hot columns. Standardized algorithms perform statistical tests on the monitoring elements and automate the data certification processes. Additionally, the DQM framework enables visualization of monitoring results, retrieval of DQM quantities from the conditions database, and dataset selections for physics analyses.

\section{Radioactive backgrounds}
\label{sec:backgrounds}

This section describes our effort to control and reduce the radioactive background in the LBC. Radio-assay analyses were performed for materials used in the detector -- particularly those installed inside the cryostat. Results of our own or public radio-assay measurements have been utilized in dedicated Monte Carlo simulations of the whole detector and served us to guide the development and construction phases of the LBC. We present event rates measured with two setups in the LBC and compare them with simulations. The original configuration of the LBC, called Setup\,1, includes two 6k$\times$4k CCDs in the CCD box with OFHC copper lids. The upgraded detector, Setup\,2, and consists of two CCD modules, each with four 6k$\times$1.5k CCDs, and EFC lids.

\subsection{Material Screening and Activation}
\label{sec:activation}
All materials used in the detector were selected either based on our radio-assay measurements or results of other experiments (DAMIC at SNOLAB\,\cite{DAMIC:2021crr}, MAJORANA\,\cite{MAJORANA:2016lsk}, SuperCDMS\,\cite{SuperCDMS:2016wui}, EDELWEISS\,\cite{EDELWEISS:2013wrh}, etc.). Our material samples have been assayed by various techniques, e.g. HPGe detectors, alpha particle counting with a XIA instrument and inductively coupled plasma mass spectrometry (ICP-MS). Results are summarized in Table\,\ref{tab:assays}. A few measurements, e.g., $^{210}$Pb in ancient Roman lead, have not been finalized yet.

\begin{table}[h]
\centering
\begin{tabular}{l|llllll}
\hline
& $^{238}$U & $^{232}$Th & $^{226}$Ra & $^{210}$Pb & $^{40}$K  \\
\hline
CCD flex & 32.1$\pm$0.5 & 1.06$\pm$0.05 &  &  & 43.9$\pm$12.9 \\
OFHC copper & 0.017$\pm$0.002 & <0.014 &  & 28.4$\pm$8.5 & <0.2 \\
EFC & 0.0013$\pm$0.0002 & 0.0014$\pm$0.0002 &  &  & 0.013$\pm$0.005 \\
Brass &  & <7.9 & <4.8 &  & <14.6 \\
Ancient Pb & <4.18 & <0.058 & <0.026 &  & <0.265 \\
TFA Pb & <60.36 & <2.72 & 0.95$\pm$1.21 & <43780 & 5.72$\pm$11.69 \\
FA Pb & <135 & <9.70 & <5.66 & 54300$\pm$2000 & <20.0 \\
PCTFE & <0.062 & <0.041 &  &  & <15.5 \\
\hline
\end{tabular}
\caption{Assay results for different isotopes in materials used in the LBC. The activity is in units of [mBq/kg]. The results for the CCD flex correspond to DAMIC-M CCD [Com.] in Tab. 4 in Ref.\,\cite{Arnquist:2023gtq}.}
\label{tab:assays}
\end{table}

We have taken steps to control the cosmogenic activation of copper and silicon parts. Storage and processing times on the surface were tracked, while also avoiding storage on the surface and transport by air when possible. However, no expedited fabrication was implemented, and no special shielded storage or transport boxes were used for the CCDs nor OFHC copper in Setup\,1 originally installed in the LBC. The surface equivalent exposure of the 6k$\times$4k CCDs was similar to the CCDs installed in DAMIC at SNOLAB\,\cite{DAMIC:2021crr}.

The 6k$\times$1.5k CCDs used in the CCD modules in Setup\,2 were fabricated from the same batch of silicon wafers which is being used for CCDs in the final DAMIC-M project. Expedited fabrication, special shielding and transport precautions were implemented before the delivery of wafers to SNOLAB, where they were stored before CCD production by Teledyne DALSA in Canada. However, no special shielding -- which has been used for CCDs in the final DAMIC-M detector -- was used during CCD production and the devices were allowed to be transported by air. Therefore, the 6k$\times$1.5k CCDs installed in the LBC have five times lower cosmogenic activation than original 6k$\times$4k CCDs. Based on work in Ref.\,\cite{Saldanha:2020ubf}, we expect $^{3}\rm{H}$ decays to contribute $\sim$2.6 and $\sim$0.5\,dru between 1 and 5\,keV in the 6k$\times$4k and 6k$\times$1.5k CCDs, respectively.

Commercial OFHC copper pieces were restricted to only ground transport and were stored underground after machining. The total exposure time of OFHC copper was approximately five months, including manufacturing and machining, before the copper storage at LSM starting in May 2021. The cosmogenic activation was greatly reduced for EFC lids grown underground at LSC in Spain and installed in the LBC in September 2022. The machining of these pieces, which had to be performed on the surface, was limited to only few days. The total exposure time, including the transport from LSC to LSM with a half-day stop for chemical cleaning, was 3.1 and 6.5 days for the top and bottom lid, respectively.

\subsection{Material Cleaning Procedures}
Cleaning procedures help remove radioactive isotopes that accumulate on material surfaces during production, storage and handling. All copper and lead inside the cryostat, and the cryostat itself, went through cleaning. We followed procedures developed at Pacific Northwest National Laboratory (PNNL) for copper\,\cite{Hoppe2007,Hoppe:2008cu} and lead\,\cite{MAJORANA:2016lsk}, which had been successfully used in other experiments. Both methods rely upon ultrasonic cleaning for gross fragment removal, followed by a  bath in an etching solution. In the case of copper (lead), pieces are submerged in an acidified-peroxide (nitric acid-hydrogen peroxide) bath to remove fine fragments, surface contamination and oxide layer. A surface passivation in a 1\% aqueous solution of citric acid is required for copper to prevent the reformation of oxides. The last steps are rinse in water and drying. Only the highest purity chemicals (Optima grade by Fisher Chemical) and deionized water (18\,M$\Omega$-cm) were used in a cleanroom at Laboratoire de Physique Subatomique \& Cosmologie in Grenoble, France. 

All instruments, tools and other parts were cleaned, dried and packed in at least two plastic bags or layers of a plastic foil before their transport to LSM. This includes copper and lead parts after their chemical cleaning.

\subsection{Simulation}
\label{sec:simulation}
A \geant\,\cite{GEANT4:2002zbu,Allison:2006ve,Allison:2016lfl} and Python-based framework was developed by the DAMIC-M collaboration to reproduce the response of the detector to various particle interactions. The first code includes a geometry model of the whole LBC, including a pixelated sensitive detector composed of silicon and polysilicon layers to model the CCD, and inactive silicon parts\,\cite{DAMIC:2021crr, DAMIC-M:2022xtp}. Other parts in the detector model are the CCD box with CCD flex cables and screws, internal lead shield, thermal guides, cold copper, cryostat, outer lead, HDPE shield and support structure.

The Livermore physics list \,\cite{Allison:2016lfl} is used for interactions of various simulated particles ($\gamma$, $\alpha$, $e^{\pm}$, neutrons, etc.). The particles originate either internally in the LBC (radioactive decays) or from the outside world, i.e. the laboratory\footnote{The information about the gamma and neutron flux together with the radon level measured at LSM are given in Sec.\,\ref{sec:design_infrastructure}.}.

The Python code is then used to simulate the detector response, i.e. pre-processing of the \geant energy deposits, electron-hole pair creation, diffusion, pixelation, dark current, readout noise, pixel saturation, and cluster finder (or clustering). This code is used to reconstruct not only simulated but also measured data.

The simulations have been invaluable in the design validation and optimization, and they have provided guidance in the selection of material type and shape. Results of the simulations for the two installed setups are summarized in Table\,\ref{tab:simulations}.
The two setups differ in the CCDs and copper lids of the CCD box, while other detector parts are the same.
The rates are provided as the average for the top and bottom CCDs between 2 and 6\,keV -- to avoid silicon and copper fluorescence X-ray lines -- in units of dru.
We verified that the contribution from $^{210}\rm{Pb}$ deposited on the outside of the CCD box and the cryostat is negligible.
The simulations do not include external gammas and neutrons -- originating either from rocks at LSM or induced by muons in the external lead. 

As can be seen in Table\,\ref{tab:simulations}, major contributors to the background rate at low energies are CCDs and the CCD box. Activities used for CCDs are based on the background model of DAMIC at SNOLAB\,\cite{DAMIC:2021crr} with two modifications. First, we rescaled the activity of cosmogenically activated isotopes in the silicon bulk ($^{3}$H, $^{22}$Na) with our knowledge of the activation history, see Sec.\,\ref{sec:activation}. Second, we estimate the activity of $^{210}$Pb on surfaces by comparing measured alpha rates in the LBC and DAMIC at SNOLAB. This contribution is 1.3 and 1.0\,dru in Setup 1 and 2, respectively. In Setup 2, $^{210}$Pb deposited on the surface of a pitch adapter gives an additional 0.7\, dru.

From Setup\,1 to Setup\,2, we have reduced the total background by 5.5\,dru according to simulations. The most significant reduction is from upgrading the lids in the CCD box to EFC. Another major decrease comes from exchanging 6k$\times$4k CCDs with two CCD modules. The predicted decrease can be compared with measurements shown in the last row in Table~\ref{tab:simulations} and discussed in Sec.\,\ref{sec:bkg_measurement}.

\begin{table}[h]
\centering
\begin{tabular}{l|c|c|c}
\hline
& Setup 1 & Setup 2 & Difference \\
\hline
CCDs        & 4.65 & 2.74 & 1.91 \\
CCD box  & 5.48 & 1.93 & 3.55 \\
Copper      & 0.06 & 0.06 & 0.0 \\
Internal lead & <0.29 & <0.29 & 0.0 \\
External parts & $\sim$0.1 & $\sim$0.1 & 0.0 \\
\hline
Total       & 10.58 & 5.12 & 5.46 \\
\hline
Measured & 12.5$\pm$2.8 & 6.7$\pm$1.1 & 5.8 \\
\hline
\end{tabular}
\caption{The background rates in dru (events/kg/kev/day) for two setups used in the LBC. Provided rates are the mean calculated for the top and bottom CCDs. In simulations, cold copper includes all copper parts but the CCD box. External gammas and neutrons were not included in simulations. The background measurements are described in Sec.\,\ref{sec:bkg_measurement}.}
\label{tab:simulations}
\end{table}

\subsection{Background measurements}
\label{sec:bkg_measurement}

Several runs dedicated to the study of the radioactive background were performed in 2022 and 2023. To measure the background, pixel binning\footnote{Pixel binning is an operating mode of the CCD where the charge of several pixels is summed before being read out. The charge can be summed only in the vertical or horizontal direction, or in both directions.} was not used for the acquired images. Instead, the entire CCD was read out after a long exposure (up to half a day). This approach was chosen to be able to effectively utilize the coincidence method for measuring bulk radioactive contamination, benefiting from the excellent spatial resolution of the CCDs.

We first verified that the external shield significantly reduces the radioactive background level. The data shows 50 times lower background rate after closing the external shield. It is important to note, that in any configuration of the external shielding, the external radiation is still attenuated by the internal lead shield inside the cryostat.

The next step was to compare the background rate measured in the original configuration -- Setup\,1 -- and after the installation of Setup\,2. The data presented in this paper have a total exposure of 0.32 and 1.2\,kg-days for Setup\,1 and 2, respectively. The event rate above 1\,keV decreased from (1.12$\pm$0.06)\,ev/g/d in Setup\,1 to (0.91$\pm$0.03)\,ev/g/d in Setup\,2. The background rate between 1 and 6\,keV, after excluding the energy range of silicon K-shell lines, is (12.5$\pm$2.8) and (6.7$\pm$1.1)\,dru in Setup\,1 and 2, respectively. The measured decrease of 5.8\,dru agrees with the expectation calculated with our \geant simulations, see Table\,\ref{tab:simulations}. We can see that measured and simulated rates agree within one and half standard deviations.

The LBC results are similar to the ones obtained in DAMIC at SNOLAB, where one CCD was in a copper box made from EFC grown at PNNL enclosed between two ancient lead bricks, and the remaining CCDs were in a box from OFHC copper\,\cite{DAMIC:2021crr}. Further, we have identified contributors to the total radioactive background in the LBC, which provides important information for the background mitigation in DAMIC-M.

An analysis of $\alpha$ decay rates and spatially coincident events, such as $\beta-\beta$ for $^{32}$Si and $^{210}$Pb, is currently underway for the surface and bulk of the 6k$\times$1.5k CCDs. Our studies follow previous work done for DAMIC at SNOLAB\,\cite{Aguilar_Arevalo_2021}. Measuring contaminants within the silicon bulk is crucial for the final DAMIC-M detector, as the CCDs used in DAMIC-M will be made from the same silicon ingot as those currently installed in the LBC.

\section{Detector Stability}
\label{sec:performance_results}
Long-term detector stability is a critical requirement on dark matter searches. In this section, we illustrate the stability of the LBC with the data collected by the Slow Control system (see Sec.\,\ref{sec:sc}) and CCDs.

\subsection{Pressure and Temperature}
\label{sec:pressure_temperature}
Our CCDs operate typically in high vacuum and low temperatures, which are achieved by implementing the following procedure. After closing the cryostat, the vacuum pump is powered on and monitored to check for any air leaks. The cryocooler is turned on when pressure inside the cryostat gets below 5$\times$10$^{-4}$\,mbar and cold copper is cooled down from room temperature to 170\,K. At this temperature, we start up the CCDs by supplying bias, clock, and substrate voltages. Necessary part of the CCD startup is the erase procedure described in Sec.~\ref{sec:ccd_performance}. The CCDs are then cooled down further to their operating temperature chosen between 110 and 130\,K. To minimize temperature-induced stress on CCDs and hence, reduce leakage current, we use a cooling rate of 0.1\,K/min.

Pressure is continuously monitored only when the CCDs are turned off to avoid light emitted by two pressure gauges. The pressure at 130\,K stays at $\sim$5$\times10^{-6}$\,mbar in the cryostat and ten times lower in a vacuum tube close to the pump.

Unlike pressure gauges, temperature sensors are monitored continuously. The RTD in the cold mass table, see Fig.\,\ref{fig:ColdCopper_Cabling}, is kept at a constant temperature by using a 100\,W heater, see Sec.\,\ref{sec:instrumentation}. The reading of an RTD epoxied in a cold finger, where it is well shielded by lead plates from CCDs, is about 2\,K higher. An RTD was kept temporarily on the CCD box during the detector commissioning  and measured 4\,K higher than the cold finger. This RTD was then removed to suppress the radioactive background. The highest temperature gradient -- about 30\,K -- is between the cold head and the cold mass table and it is due to a polyimide tape, which provides an electrical isolation but reduces thermal contact. The temperature of the CCDs has been measured through independent thermal tests and is typically $\sim$5\,K higher than the temperature measured by an RTD epoxied on the outside of the CCD box.

Over 63 consecutive days of data taking (described in Refs.\,\cite{DAMIC-M:2023gxo,DAMIC-M:2023hgj}), the cold mass table  temperature increased only slightly from 116.9\,K to 117.2\,K, see Fig.\,\ref{fig:Stability}. This increase may be explained either by heating the cryostat by a flow of air warmed by the cryocooler body when the external shielding is closed, or relaxing the connection between cold copper parts. This slow temperature increase was no issue during the data taking, because it did not cause any change of measured dark current as is shown in Refs.\,\cite{DAMIC-M:2023hgj}.

\subsection{CCD Performance}
\label{sec:ccd_performance}
The LBC serves as a test setup for the DAMIC-M project and allows us to perform various studies of operating parameters for different CCD formats and electronics. An extensive optimization program for clock and bias voltages, and timings were carried out. These tests were performed with the originally installed 6k$\times$4k CCDs and then the CCD modules with 6k$\times$1.5k devices. The latter is of particular importance, because very similar CCD modules will be installed in the full-scale DAMIC-M detector.

Details on the image processing and selection criteria are described in Ref.\,\cite{DAMIC-M:2022xtp,DAMIC-M:2023gxo}. The readout noise $\sigma_{1}$ can be obtained from single-skip images, but multi-skip images ($N_{\rm{skip}}\gg1$) are necessary to measure single electron peaks. By fitting the pedestal (0\,e$^-$) and 1\,e$^-$ peaks in the pixel charge distribution with a multi-Gaussian function\footnote{The multi-Gaussian function is composed of equally spaced Gaussian functions.} convolved with a Poisson distribution, we obtain the readout noise, gain (or calibration constant), and dark current. These parameters have been extensively studied with all setups installed in the LBC.

The first studied noise levels were $\sigma_{1}\sim10$\,e$^-$ which are higher than in DAMIC at SNOLAB and test setups located in different institutions. After implementing noise mitigation steps (e.g. improved electrical grounding and better timing of voltage clocks), the noise level decreased to an acceptable level of $\sigma_{1}\sim7$\,e$^-$ in images taken with ARC CCD controllers in Setup 1. The averaging in the skipper amplifier reduces any remaining high frequency noise contributions.

When reading CCD modules with two unsynchronized ARC CCD controllers in Setup 2, a significant correlated noise was observed. Because this will not be the case with the new electronics developed for DAMIC-M, we have not invested resources to synchronize the controllers. Reading CCD modules separately and applying a noise decorrelation algorithm, provided the resolution at the same level as when we read a single CCD.

The major noise reduction was achieved with the new electronics developed for DAMIC-M. Noise measured by two CCD modules run synchronously is $\sigma_{1}\leq 3$\,e$^-$\,rms. When we read the CCDs with $N_{\rm{skip}}=500$, the skipper averaging reduces noise to $\sim0.1$\,e$^-$.

Various clock and bias voltages were investigated to study the clock induced charge, amplifier glowing, and charge transfer inefficiency, all three of which have an important effect on the e$^{-}$ rate. During these studies, we have employed also a  $^{60}$Co radioactive source. In low dark current runs, we have been using rather low substrate bias voltage (V$_\mathrm{sub}=45$\,V) and swing of vertical and horizontal clocks (between 3 and 3.5\,V). These parameters have been modified in other studies, e.g. in background runs.

To achieve a low dark current, the CCD is started up at high temperature (170\,K) and an erase procedure is performed as suggested by Ref.\,\cite{Holland:2003zz}. Our erase procedure consists of the following steps: reduce V$_\mathrm{sub}$ to 0\,V in a linear way ($\sim$15\,V/s) and put clocks in inversion phase (i.e. +9\,V). After a few seconds, we increase V$_\mathrm{sub}$ back to its normal operating value and restore the clocks to their normal values for charge transfer at V$_\mathrm{sub}$=10\,V. Temperature is then reduced to its target value.

\begin{figure*}[!t]
    \centering
    \includegraphics[width=0.9\textwidth, trim=0 0 0 0, clip=true]{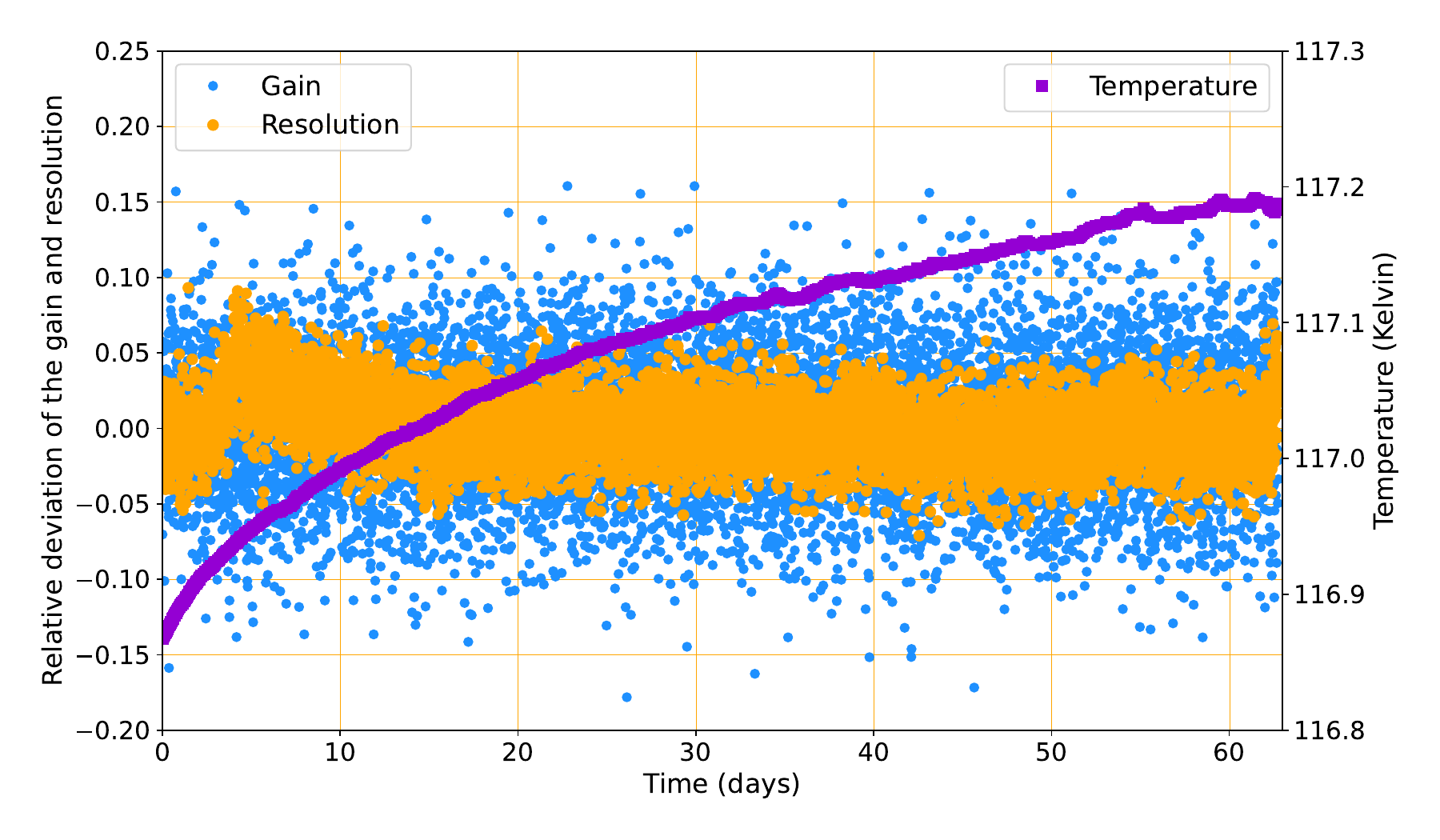}
    \caption{The relative deviation of the gain (blue) and resolution (yellow) of the U amplifier in CCD 6414 during 63-day run. Temperature of cold copper (dark violet) in Kelvins is shown on the right y-axis. See text for more details.}
    \label{fig:Stability}
\end{figure*}

The stability of the resolution and gain measured by an amplifier in one CCD during a 63-day long continuous run in the summer 2022 is shown in Fig.\,\ref{fig:Stability}. The corresponding dark current can be found in Fig.\,1 in Ref.\,\cite{DAMIC-M:2023hgj}. Even though the temperature of cold copper increased by 0.3\,K over this period, we did not observe any change of either the amplifier gain or the resolution. The gain and resolution of the amplifier U in CCD 6414 had standard deviation of 4.4\% and 2.1\%, respectively.
Due to the excellent stability of the LBC, we could constrain DM-electron scattering by exploiting the expected daily modulation due to DM interactions within the Earth\,\cite{DAMIC-M:2023hgj}.

We performed additional tests to characterize installed skipper CCDs. These include study the temperature dependence of dark current and the increase in dark current for different levels of ionizing radiation from the external $^{60}$Co source. The analysis is ongoing and will be published elsewhere.

\section{Conclusions}
The DAMIC-M prototype detector, the LBC, was built at LSM in late 2021 and continues to serve as a low background setup for skipper CCDs. The LBC has operated without any major issues over its lifetime and its stable operation has lead to the best published limits on the interaction of DM with electrons\,\cite{DAMIC-M:2023gxo,DAMIC-M:2023hgj}. Of critical importance, we have studied the CCD performance in a low background environment and after numerous thermal cycles. Thanks to successful background mitigation steps, the background level was reduced to (6.7$\pm$1.1)\,dru between 1 and 6\, keV. This result, together with detailed simulations, provides a clear pathway for further background reduction in the final DAMIC-M detector, which will be built at LSM in 2025.

\section{Acknowledgements}
The DAMIC-M project has received funding from the European Research Council (ERC) under the European Union's Horizon 2020 research and innovation programme Grant Agreement No. 788137, and from NSF through Grant No. NSF PHY-1812654. 
The work at University of Chicago and University of Washington was supported through Grant No. NSF PHY-2110585.
This work was supported by the Kavli Institute for Cosmological Physics at the University of Chicago through an endowment from the Kavli Foundation. 
We thank the College of Arts and Sciences at UW for contributing the first CCDs to the DAMIC-M project.
Part of this work was conducted at the Washington Nanofabrication Facility / Molecular Analysis Facility, a National Nanotechnology Coordinated Infrastructure (NNCI) site at the University of Washington with partial support from the National Science Foundation via awards NNCI-1542101 and NNCI-2025489.
We also thank the Krieger School of Arts \& Science at Johns Hopkins University for its contributions to the DAMIC-M experiment.
IFCA was supported by project PID2019-109829GB-I00 funded by MCIN/ AEI /10.13039/501100011033.
The Centro At\'{o}mico Bariloche group is supported by ANPCyT grant PICT-2018-03069.
The University of Z\"{u}rich was supported by the Swiss National Science Foundation.
The CCD development work at Lawrence Berkeley National Laboratory MicroSystems Lab was supported in part by the Director, Office of Science, of the U.S. Department of Energy under Contract No. DE-AC02-05CH11231.

We would like to thank the Modane Underground Laboratory and its staff for support through underground space, logistical and technical services. LSM operations are supported by the CNRS, with underground access facilitated by the Societe Francaise du Tunnel Routier du Frejus

\bibliographystyle{style/h-physrev}
\bibliography{literature.bib}

\end{document}

%% file: 0_authorlist.tex
\collaboration{DAMIC-M Collaboration}

\author[a]{I.~Arnquist,}
\author[b]{N.~Avalos,}
\author[c]{P.~Bailly,}
\author[d,*]{D.~Baxter,}
\author[b]{X.~Bertou,}
\author[d]{M.~Bogdan,}
\author[e]{C.~Bourgeois,}
\author[d]{J.~Brandt,}
\author[f]{A.~Cadiou,}
\author[g]{N.~Castell\'{o}-Mor,}
\author[h]{A.E.~Chavarria,}
\author[h]{M.~Conde,}
\author[d]{J.~Cuevas-Zepeda,}
\author[i]{A.~Dastgheibi-Fard,}
\author[c]{C.~De Dominicis,}
\author[e]{O.~Deligny,}
\author[d]{R.~Desani,}
\author[c]{M.~Dhellot,}
\author[g]{J.~Duarte-Campderros,}
\author[b]{E.~Estrada,}
\author[j]{D.~Florin,}
\author[j]{N.~Gadola,}
\author[c]{R.~Ga\"{i}or,}
\author[j]{E.-L.~Gkougkousis,}
\author[g]{J.~Gonz\'{a}lez S\'{a}nchez,}
\author[l]{S.~Hope,}
\author[a]{T.~Hossbach,}
\author[h]{M.~Huehn,}
\author[h]{M.~Kallander,}
\author[j]{B.~Kilminster,}
\author[c]{L.~Iddir,}
\author[g]{A.~Lantero-Barreda,}
\author[k]{I.~Lawson,}
\author[c]{H.~Lebbolo,}
\author[j]{S.~Lee,}
\author[f]{P.~Leray,}
\author[c]{A.~Letessier Selvon,}
\author[h]{H.~Lin,}
\author[e]{P.~Loaiza,}
\author[g]{A.~Lopez-Virto,}
\author[c]{D.~Martin,}
\author[h]{K.J.~McGuire,}
\author[f]{T.~Milleto,}
\author[h]{P.~Mitra,}
\author[g]{D.~Moya Martin,}
\author[d]{S.~Munagavalasa,}
\author[l]{D.~Norcini,}
\author[a]{C.~Overman,}
\author[c]{S.~Paul,}
\author[h]{D.~Peterson,}
\author[h]{A.~Piers,}
\author[e]{O.~Pochon,}
\author[c,d]{P.~Privitera,}
\author[e]{D.~Reynet,}
\author[d]{B.A.~Roach,}
\author[j]{P.~Robmann,}
\author[h]{R.~Roehnelt,}
\author[f]{M.~Settimo,}
\author[l]{S.~Smee,}
\author[d]{R.~Smida,}
\author[d]{B.~Stillwell,}
\author[h]{T.~Van~Wechel,}
\author[c,h]{M.~Traina,}
\author[g]{R.~Vilar,}
\author[j]{A.~Vollhardt,}
\author[i]{G.~Warot,}
\author[j]{D.~Wolf,}
\author[d]{R.~Yajur,}
\author[c]{and~J-P.~Zopounidis}

\affiliation[a]{Pacific Northwest National Laboratory (PNNL), Richland, WA, United States}

\affiliation[b]{Centro At\'{o}mico Bariloche and Instituto Balseiro, Comisi\'{o}n Nacional de Energ\'{i}a At\'{o}mica (CNEA), Consejo Nacional de Investigaciones Cient\'{i}ficas y T\'{e}cnicas (CONICET), Universidad Nacional de Cuyo (UNCUYO), San Carlos de Bariloche, Argentina}

\affiliation[c]{Laboratoire de physique nucl\'{e}aire et des hautes \'{e}nergies (LPNHE), Sorbonne Universit\'{e}, Universit\'{e} Paris Cit\'{e}, CNRS/IN2P3, Paris, France}

\affiliation[d]{Kavli Institute for Cosmological Physics and The Enrico Fermi Institute, The University of Chicago, Chicago, IL, United States}

\affiliation[e]{IJCLab, CNRS/IN2P3, Universit\'{e} Paris-Saclay, Orsay, France}

\affiliation[f]{SUBATECH, Nantes Universit\'{e}, IMT Atlantique, CNRS-IN2P3, Nantes, France}

\affiliation[g]{Instituto de F\'{i}sica de Cantabria (IFCA), CSIC - Universidad de Cantabria, Santander, Spain}

\affiliation[h]{Center for Experimental Nuclear Physics and Astrophysics, University of Washington, Seattle, WA, United States}

\affiliation[i]{LPSC LSM, CNRS/IN2P3, Universit\'{e} Grenoble-Alpes, Grenoble, France}

\affiliation[j]{Universit\"{a}t Z\"{u}rich Physik Institut, Z\"{u}rich, Switzerland}

\affiliation[k]{SNOLAB, Lively, ON, Canada}

\affiliation[l]{Department of Physics and Astronomy, Johns Hopkins University, Baltimore, MD , USA}

\affiliation[*]{Now at Fermi National Accelerator Laboratory, Batavia, IL, United States}

\emailAdd{damicmcollaboration@gmail.com}